\documentclass[runningheads]{llncs}
\usepackage{cite}
\usepackage{authblk}
\usepackage{subcaption}
\usepackage{amsmath,amssymb,amsfonts}
\usepackage{graphicx}
\usepackage{textcomp}
\usepackage{xcolor}
\usepackage[english]{babel}
\usepackage[utf8]{inputenc}
\usepackage{hyperref}
\usepackage[utf8]{inputenc}
\usepackage[english]{babel}

\usepackage[font={small}]{caption}

\usepackage[linesnumbered,ruled,vlined]{algorithm2e}

\SetCommentSty{mycommfont}

\SetKwInput{KwInput}{Input}                % Set the Input
\SetKwInput{KwOutput}{Output}              % set the Output

\begin{document}
\title{Online Memory Leak Detection in the Cloud-based Infrastructures}
\author{Anshul Jindal\inst{1}\orcidID{0000-0002-7773-5342} \and
Paul Staab\inst{2} \and
Jorge Cardoso\inst{2}\orcidID{0000-0001-8992-3466} \and
Michael Gerndt\inst{1}\orcidID{0000-0002-3210-5048} \and
Vladimir Podolskiy\inst{1}\orcidID{0000-0002-2775-3630}}
\authorrunning{Jindal et al.}
\institute{Chair of Computer Architecture and Parallel Systems, \\Technical University of Munich, Garching, Germany  \\
\email{{anshul.jindal@tum.de, gerndt@in.tum.de, v.podolskiy@tum.de}}\\ \and
Huawei Munich Research Center, Huawei Technologies Munich,Germany\\
\email{\{\}@huawei.com}}
\maketitle              % typeset the header of the contribution
\begin{abstract}
A memory leak in an application deployed on the cloud can affect the availability and reliability of the application. Therefore, to identify and ultimately resolve it quickly is highly important. However,  in the production environment running on the cloud, memory leak detection is a challenge without the knowledge of the application or its internal object allocation details.

This paper addresses this \textit{challenge of online detection of memory leaks in cloud-based infrastructure without having any internal application knowledge} by introducing a novel machine learning based algorithm Precog. This algorithm solely uses one metric i.e the system's memory utilization on which the application is deployed for the detection of a memory leak. The developed algorithm's accuracy was tested on 60 virtual machines manually labeled memory utilization data provided by our industry partner Huawei Munich Research Center and it was found that the proposed algorithm achieves the accuracy score of 85\%  with less than half a second prediction time per virtual machine. %The same algorithm also achieves this by decreasing the overall compute time by 80\% when compared to a developed baseline algorithm called Linear Backward Regression.

\keywords{ memory leak \and  online memory leak detection \and  memory leak patterns \and  cloud\and  linear regression}
\end{abstract}
\section{Introduction}
Cloud computing is widely used in the industries for its capability to provide cheap and on-demand access to compute and storage resources. Physical servers resources located at different data centers are split among the virtual machines (VMs) hosted on it and distributed to the users~\cite{7570950}. Users can then deploy their applications on these VMs with only the required resources. This allows the efficient usage of the physical hardware and reducing the overall cost. However, with all the advantages of cloud computing there exists the drawback of detecting a fault or an error in an application or in a VM efficiently due to the layered virtualisation stack~\cite{7416355, 7977362}. A small fault somewhere in the system can impact the performance of the application. 

An application when deployed on a VM usually requires different system resources such as memory, CPU and network for the completion of a task. If an application is mostly using the memory for the processing of the tasks then this application is called a memory-intensive application~\cite{6799629}. It is the responsibility of the application to release the system resources when they are no longer needed. When such an application fails to release the memory resources, a \textbf{memory leak} occurs in the application~\cite{Xie:2005:CPM:1095430.1081728}. Memory leak issues in the application can cause continuous blocking of the VM's resources which may in turn result in slower response times or application failure. In software industry, memory leaks are treated with utmost seriousness and priority as the impact of a memory leak could be catastrophic to the whole system. In the development environment, these issues are rather easily detectable with the help of static source code analysis tools or by analyzing the heap dumps. But in the production environment running on the cloud, memory leak detection is a challenge and it only gets detected when there is an abnormality in the run time, abnormal usage of the system resources, crash of the application or restart of the VM. Then the resolution of such an issue is done at the cost of compromising the availability and reliability of the application. Therefore it is necessary to monitor every application for memory leak and have an automatic detection mechanism for memory leak before it actually occurs. However, it is a challenge to detect memory leak of an application running on a VM in the cloud without the knowledge of the programming language of the application, nor the knowledge of source code nor the low level details such as allocation times of objects, object staleness, or the object references~\cite{Sor2011ASA}. Due to the low down time requirements for the applications running on the cloud, detection of issues and their resolutions is to be done as quickly as possible. Therefore, this challenge is addressed in this paper  \textit{ by solely using the VM's memory utilization as the main metric and devising a novel algorithm called \textbf{Precog} to detect memory leak}. %Therefore, this point is also kept into consideration while creating the memory leak detection algorithms.

The main contribution of this paper are as follows: 
\begin{itemize}
    \item \textbf{Algorithm}: We propose an online novel machine learning based algorithm \textbf{Precog} for accurate and efficient detection of memory leaks by solely using the VM's memory utilization as the main metric.
    
    \item \textbf{Effectiveness}: Our proposed algorithm achieves the accuracy score of 85\% on the evaluated dataset provided by our industry partner and accuracy score of above 90\% on the synthetic data generated by us. 
    
    \item \textbf{Scalability}: Precog's predict functionality is linearly scalable with the number of values and takes less than a second for predicting in a timeseries with 100,000 values. 
\end{itemize}
\textbf{Reproducibility}: our code and synthetic generated data are publicly available at:  \url{https://github.com/ansjin/memory\_leak\_detection}.

\section{Related Work}
%Automatic garbage collection is one of the most important aspects of modern day programming languages, which simplifies the life of application developers taking away the burden of manual memory management. Programming languages such as \textbf{C} and \textbf{C++}, memory related faults occur due one of the following reasons \textit{1. Dangling pointers} - de-referencing pointers to objects that the program previously freed, \textit{2 lost pointers} – losing all pointers to objects that the program neglects to free, and \textit{(3) unnecessary references} – keeping pointers to objects the program never uses again. In garbage collected programming languages, GC algorithms are designed to automatically deal with \textit{1 and 2}. However, keeping a reference to an unused object can not be tracked by garbage collection and program starts leaking memory. Thus, a memory leak in a garbage-collected language occurs when a program continue to keep the references to objects that are no longer needed, preventing the garbage collector from reclaiming space. Best in such cases, the object is not a growing instance, and results in constant amount of memory being leaked. But in worst cases, a growing object with unused parts will cause the program/applications to run out of memory and crash~\cite{Jump:2007:CDM:1190216.1190224}. These type of memory fault defects are very difficult to debug and fix given the critical nature of the defect as the ultimate object that crashes the application may not necessarily be of leaking nature. 

Memory leak detection has been studied over the years and several solutions have been proposed. Sor et al. reviewed different memory leak detection approaches based on their implementation complexity, measured metrics, and intrusiveness and a classification taxonomy was proposed~\cite{Sor2014MemoryLD}. The classification taxonomy broadly divided the detection algorithms into \textit{(1) Online detection, (2) Offline detection and (3) Hybrid detection}. The \textit{online detection} category uses either staleness measure of the allocated objects or their growth analysis. \textit{Offline detection} category includes the algorithms that make use of captured states i.e heap dumps or use a visualization mechanism to manually detect memory leaks or use static source code analysis. \textit{Hybrid detection} category methods combine the features offered by online and offline methods to detect memory leaks. Our work falls in the category of online detection therefore, we now restrict our discussion to the approaches related to the online detection category only. 

Based on the staleness measure of allocated objects, Rudaf et al. proposed "LeakSpot" for detecting memory leaks in web applications~\cite{Rudafshani:2017:LDD:3035064.3035070}. It locates JavaScript allocation and reference sites that produce and retain increasing numbers of objects over time and uses staleness as a heuristic to identify memory leaks. Vladimir \v{S}or et al. proposes a statistical metric called \textit{genCount} for memory leak detection in Java applications~\cite{Sor2015MemoryLD}. It uses the number of different generations of the objects grouped by their allocation sites, to abstract the object staleness - an important attribute indicating a memory leak.  Vilk et al. proposed a browser leak debugger for automatically debugging memory leaks in web applications called as "BLeak"~\cite{Vilk:2018:BAD:3192366.3192376}. It collects heap snapshots and analyzes these snapshots over time to identify and rank leaks. BLeak targets application source code when detecting memory leaks. 

Based on the growth analysis objects, Jump et al. proposes "Cork" which finds the growth of heap data structure via a directed graph  \textit{Type Points-From Graph} - TPFG, a data structure which describes an object and its outgoing reference~\cite{Jump:2007:CDM:1190216.1190224}. To find memory leaks, TPFG's growth is analyzed over time in terms of growing types such as a list. FindLeaks proposed by Chen et al. tracks object creation and destruction and if more objects are created than destroyed per class then the memory leak is found~\cite{4291098}. Nick Mitchell and Gary Sevitsky proposed "LeakBot", which looks for heap size growth patterns in the heap graphs of Java applications to find memory leaks~\cite{10.1007/978-3-540-45070-2_16}. "LEAKPOINT" proposed by Clause et al. uses dynamic tainting to track heap memory pointers and further analyze it to detect memory leaks~\cite{6062055}.  

Most of the online detection algorithms that are proposed focus either on the programming language of the running application or on garbage collection strategies or the internals of the application based on the object's allocation, references, and deallocation. To the best of our knowledge, there is no previous work that solely focuses on the detection of memory leaks using just the system's memory utilization data on which application is deployed. The work in this paper, therefore, focuses on the detection of a memory leak pattern irrespective of the programming language of the application or the knowledge of application's source code or the low-level details such as allocation times of objects, object staleness, or the object references. 
\section{Methodology for Memory Leak Detection}
In this section, we present the problem statement of memory leak detection and describes our proposed algorithm's workflow for solving it. 
\subsection{Problem Statement}
Table~\ref{tab1:symbols} shows the symbols used in this paper.
\begin{table}[t]
\caption{Symbols and definitions.}\label{tab1:symbols}
\begin{tabular*}{\textwidth}{l @{\hskip 0.1in} l}
\hline
\textbf{Symbol} &  \textbf{Interpretation}\\
\hline
$t$ & a timestamp \\
$x_t$ &  the percentage utilization of a resource (for example memory  \\
 & or disk usage) of a virtual machine at time $t$\\

$N$ & Number of data points \\
$x = \{x_1, x_2, ..., x_N\}$ & a VM's memory utilization observations from the Cloud\\
$T$ & time series window length\\
$x_{t - T:t}$ & a sequence of observations $\{x_{t - T}, x_{t - T+1}, ..., x_t\}$ from \\
& time ${t - T}$ to $t$\\

$U$ & percentage memory utilization threshold equal to 100.\\
$C$ & critical time\\
\hline
\end{tabular*}
\end{table}

We are given $x = \{x_1, x_2, ..., x_N\}$, an $N × 1$ dataset representing the memory utilization observations of the VM and an observation $x_t \in R$ is the percentage memory utilization of a virtual machine at time $t$.  The objective of this work is to determine whether or not there is a memory leak on a VM such that an observation $x_t$ at time $t$ reaches the threshold $U$ memory utilization following a trend in the defined critical time $C$. Formally:

\begin{problem}{ (Memory Leak Detection) }

\begin{itemize}
    \item \textbf{Given}: a univariate dataset of $N$ time ticks, $x = \{x_1, x_2, ..., x_N\}$, representing the memory utilization observations of the VM.
    
    \item \textbf{Output}: an anomalous window for a VM consisting of a sequence of observations $x_{t - T:t}$ such that these observations after following a certain trend will reach the threshold $U$ memory utilization at time $t+M$ where $M \leq C$.
    
\end{itemize}
\end{problem}

\begin{definition}{ (Critical Time) } It is the maximum time considered relevant for reporting a memory leak in which if the trend line of memory utilization of VM is projected, it will reach the threshold $U$.  

\end{definition}

%A time series data contains successive observations which are usually collected at equal-space timestamps. In our study, we focus on
%a virtual machine data from the Cloud, defined as $x = \{x_1, x_2, ..., x_N\}$, where $N$ is the length of $x$ and an observation $x_t \in R$ is the percentage utilization of a resource (for example memory or disk usage) of a virtual machine at time $t$ ($t \leq N$). A time series window of length $T$ referred by $x_{t - T:t}$ ($\in R_{(T +1}$) is used to denote a sequence of observations $\{x_{t - T}, x_{t - T+1}, ..., x_t\}$ from time ${t - T}$ to $t$.

%The proposed approach can be applied for multiple VMs as well. %returns an anomalous window for a VM consisting of a sequence of observations $x_{t - T:t}$ such that these observations after following a certain trend will reach the threshold $U$ memory utilization at time $t+M$ where $M \leq C$. The same approach can be applied for multiple VMs as well. 

\subsection{Illustrative Example}
\begin{figure}[htbp]
\centerline{\includegraphics[width=0.9\linewidth]{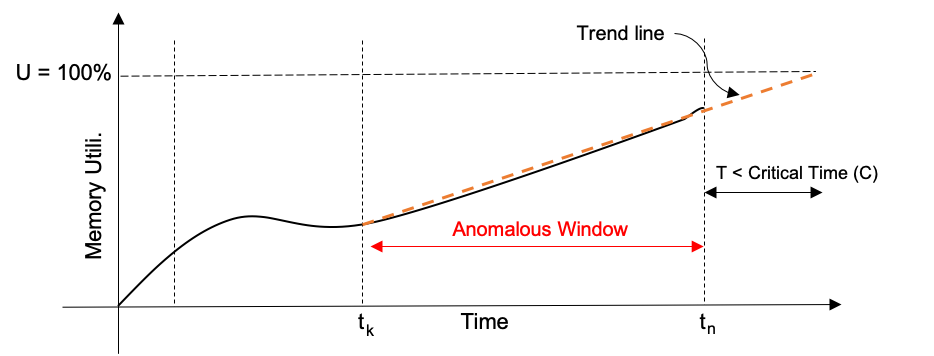}}
\caption{Example memory utilization of a memory leaking VM with the marked anomalous window.}
\label{fig_mem_leak_detection_problem}
\end{figure}

Fig.~\ref{fig_mem_leak_detection_problem} shows the example memory utilization of a memory leaking VM with the marked anomalous window between $t_k$ and $t_n$. It shows that the memory utilization of the VM will reach the defined threshold ($U = 100\%$) within the defined critical time $C$ by following a linearly increasing trend (shown by the trend line) from the observations in the anomalous window. Therefore, this VM is regarded as a memory leaking VM.

%The current work focuses on the memory leak detection, therefore $x = \{x_1, x_2, ..., x_N\}$ represents the memory utilization observations of the VM. 
%The objective of this work is to determine whether or not there is a memory leak on a VM such that an observation $x_t$ at time $t$ reaches the threshold $U$ ($U \leq 100$) memory utilization following a trend in the defined critical time $C$. 
%\begin{definition}{ (Exit Time) } The time required to reach the %defined threshold. 
%\end{definition}

Our developed approach  can be applied for multiple VMs as well. We also have conducted an experiment to understand the memory usage patterns of memory leak applications. We found that, if an application has a memory leak, usually the memory usage of the VM on which it is running increases steadily. It continues to do so until all the available memory of the system is exhausted. This usually causes the application attempting to allocate the memory to terminate itself. Thus, usually a memory leak behaviour exhibits a linearly increasing or "sawtooth" memory utilization pattern.

\subsection{Memory Leak Detection Algorithm: Precog}
\begin{figure}[htbp]
\centerline{\includegraphics[width=1\linewidth]{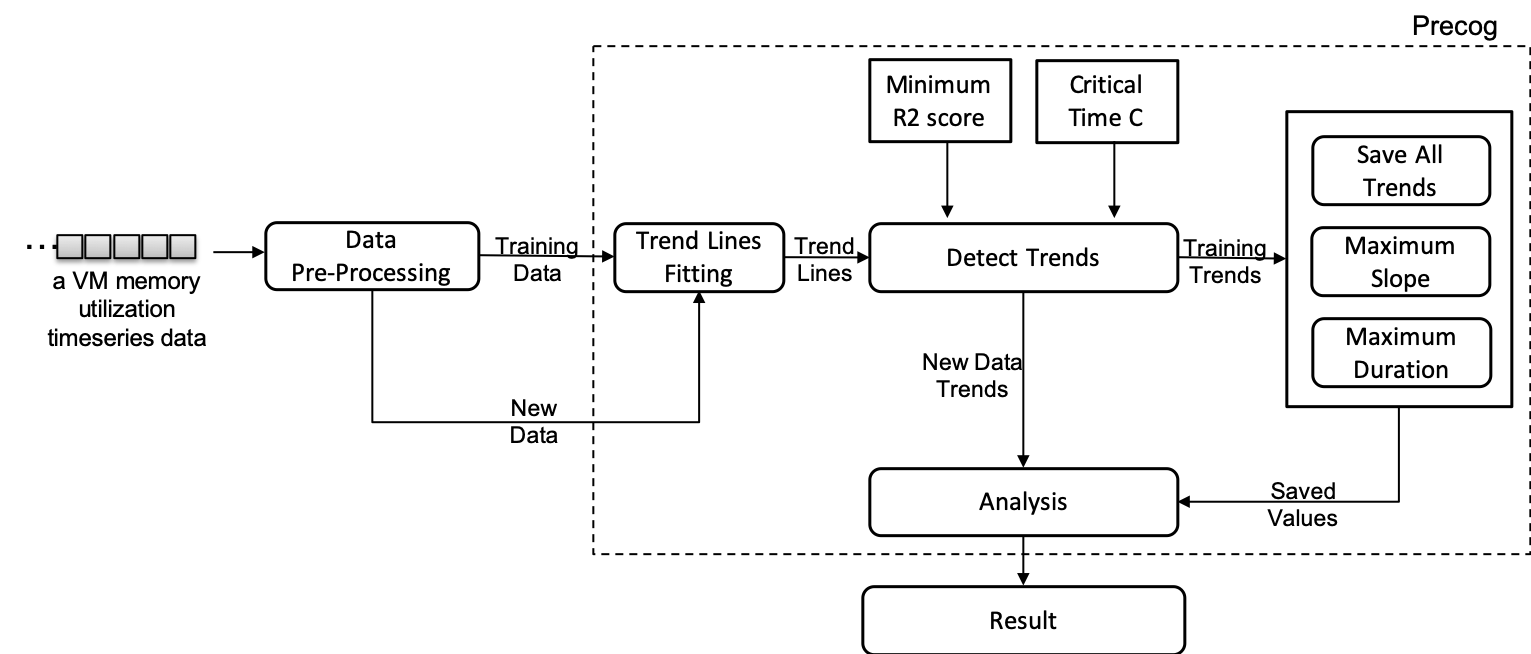}}
\caption{Overall workflow of Precog algorithm. }
\label{fig_precog}
\end{figure}

The Precog algorithm consists of two phases: offline training and online detection. Fig.~\ref{fig_precog} shows the overall workflow of the Precog algorithm.  

\textbf{Offline training}: The procedure starts by collecting the memory utilization data of a VM and passing it to \textit{Data Pre-processing} module, where the dataset is first transformed by resampling  the number of observations to one every defined resampling time resolution and then the time series data is median smoothed over the specified smoothing window. In \textit{Trend Lines Fitting} module, firstly, on the whole dataset, the change points $P = \{P_1, P_2, ..., P_k\}$, where $k \leq n-1$,  are detected. By default, two change points one at the beginning and other at the end of time series data are added. If the change points are not detected, then the algorithm will have to go though each data point and it will be compute intensive, therefore these points allows the algorithm to directly jump from one change point to another and selecting all the points in between the two change points. \textit{Trend Lines Fitting} module selects a sequence of observations $x_{t - L:t}$ between the two change points: one fixed $P_1$ and other variable $P_{r}$ where $r \leq k$  and a line is fitted on them using the linear regression. The R-squared score, size of the window called as \textit{duration}, time to reach threshold called \textit{exit time} and slope of line are calculated. This procedure is repeated with keeping the fixed change point the same and varying the other for all other change points. Out of all the fitted lines, the best-fitted line based on the largest duration and highest slope is selected for the fixed change point. If this best-fitted lines' time to reach threshold falls below the critical time then its slope and duration are saved as historic trends. 
    
This above procedure is again repeated by changing the fixed change point to all the other change points. At the end of this whole procedure, we get for each change point, a best-fitted trend if it exists. Amongst the captured trends, maximum duration and the maximum slope of the trends are also calculated and saved. This training procedure can be conducted routinely, e.g., once per day or week. The method's pseudocode is shown in the  algorithm's~\ref{alg:precog} \textit{Train function}.

\textbf{Online detection}: In the Online Detection phase, for a new set of observations $\{x_k, x_k+1, x_k+2, ...,x_k+t-1 x_k+t\}$ from time $k$ to $t$ where $t - k \geq P_{min}$ belonging to a VM  after pre-processing is fed into the \textit{Trend Lines Fitting} module.  In \textit{Trend Lines Fitting} module, the change points are detected. A sequence of observations $x_{t - L:t}$ between the last two change points starting from the end of the time series are selected and a line is fitted on them using the linear regression. The R-squared score, slope, duration and exit time to reach threshold of the fitted line is calculated. If its slope and duration are greater than the saved maximum counter parts then that window is marked anomalous. Otherwise, the values are compared against all the found training trends and if fitted-line's slope and duration are found to be greater than any of the saved trend then, again that window will be marked as anomalous. This procedure is further repeated by analyzing the observations between the last change point $P_k$ and the previous next change point until all the change points are used. This is done for the cases where the new data has a similar trend as the historic data but now with a higher slope and longer duration. The algorithm's pseudo code showing the training and test method are shown in the  algorithm~\ref{alg:precog}.

\begin{definition}{ (Change Points) } A set of time ticks  which deviate highly from the normal pattern of the data.  This is calculated by first taking the first-order difference of the input timeseries. Then, taking their absolute values and calculating their Z-scores. The indexes of observations whose Z-scores are greater than the defined threshold (3 times the standard deviation) represents the change points. The method's pseudocode is shown in the  algorithm's~\ref{alg:precog} \textit{CPD function}.
\end{definition}

%For time series modeling, historical values are useful for understanding current data. Therefore, a sequence of observations $x_{t - L:t}$ instead of just $x_t$ belonging to a VM passed from \textit{Data Pre-processing} module is used by the individual algorithm module for a potential memory leak detection.

%\textit{Data Pre-Processing} and \textit{Trend Lines Fitting} modules are shared by both the phases. 

%Metric time series data belonging to a VM  is passed by the \textit{Data Pre-Processing} module to the algorithm's \textit{Trend Lines Fitting} module. 

%In this algorithm, firstly, on the whole dataset the change points $P = \{P_1, P_2, ..., P_k\}$, where $k \leq n-1$,  are detected. 

%As in LBRCPD, the best-fitted line's R-squared score is compared with the defined minimum R-squared score $R_{min}$ and uses the best-fitted line model to calculate the time to reach the defined threshold. If the calculated R-squared score is above the defined minimum R-squared score and time to reach threshold is within the defined critical time then  

%New trends and training trends can further be combined if possible to form a bigger trend for better analysis but that part is not yet implemented as part of the algorithm.

\begin{algorithm}
\DontPrintSemicolon
  
  \KwInput{input\_Train\_Ts$, $R2\_score\_{min}, input\_Test\_Ts,  critical\_time}
  \KwOutput{anomalous list a}
  
  % Set Function Names
  \SetKwFunction{Ftrain}{TRAINING}
  \SetKwFunction{Fcpd}{CPD}
  \SetKwFunction{FTest}{TEST}
 
% Write Function with word ``Function''

    \SetKwProg{Fn}{Function}{:}{}
  \Fn{\Fcpd{$x=input\_Ts, threshold = 3$}}{ 
       $\textit{absDiffTs} = \text{first order absolute difference of }\textit{x}$\;
       $\textit{zScores} = \text{calculate z-scores of }\textit{absDiffTs}$\;
       $\textit{cpdIndexes} = \text{indexes of }\textit{(zScores} > \textit{threshold)}$  \;
       \textbf{return} $\textit{cpdIndexes}$\tcp*{return the change-points indexes}
  }

  \SetKwProg{Fn}{Function}{:}{}
  \Fn{\Ftrain{$x=input\_Train\_Ts$, $R2\_score\_{min}, C=critical\_time$}}{ 
  \tcp{Train on $\textit{input\_Train\_Ts}$}
        P = \textbf{CPD}(x) \tcp*{get Change-points}
        p1 = 0 \;
        \While{p1 $<=$ length(P)}
       {
       		p2 = p1 \;
       		${D_{b}, S_{b}, T_{b}} =  0$  \tcp*{best local trend's duration, slope, exit time} 
       		 \While{p2 $<=$ length(P)}
               {
               		$exit\_time, r2, dur, slope \gets \textit{\textbf{LinearRegression}(ts)}$ \tcp*{fitted line's exit time, R2 score,  duration, slope}
               		\If{$r2\geq R2\_score\_{min} \text{ and } dur\geq D_{b} \text{ and } slope\geq S_{b}$}
                    {
                        \textbf{Update}(${D_{b}, S_{b}, T_{b}}$) \tcp*{update best local values}
                    }
                    $p2 = p2+1$ \;
                }
            \If{$T_{b}\leq C$}
            {   \If{$D_{b}\geq D_{max} \text{ and } S_{b}\geq S_{max}$}
                    {
                        \textbf{Update}(${D_{max}, S_{max}}$) \tcp*{update global trend values}
                    }
                \textbf{saveTrend}($D_{b}, S_{b}), \textbf{ save}({D_{max}}, S_{max}$) \tcp*{save values}
            }
       	     $p1 = p1+1$	
       }
  }
   \SetKwProg{Fn}{Function}{:}{}
  \Fn{\FTest{$x=input\_Test\_Ts, C=critical\_time$}}{ 
  \tcp{Test on the new data to find anomalous memory leak window}
        ${a} = \text{[0] }$ \tcp*{anomalous empty array of size \textit{input\_Test\_Ts}}
        P = \textbf{CPD}(x) \tcp*{get Change-points}
        ${len} = \text{length(P)}$ \tcp*{length of change point indexes}
        
        \While{$i\leq len$} 
       {
       		$\textit{ts} =  x[P[len-i ] : P[len]]$  \tcp*{i is a loop variable}
       		$exit\_time, r2, dur, slope = \textit{\textbf{LinearRegression}(ts)}$ \;
       		${ D_{max}, S_{max}, Trends} = \text{get saved values }$ \;
            \If{$exit\_time,\leq C \text{ and } r2\geq R_{min} $}
            {   \If{$ slope\geq S_{max} \text{ and } dur\geq D_{max}$}
                    {
                        $a[P[len-i ] : P[len]] = 1$ \tcp*{current trend greater than global saved so mark anomalous}
                    }
                \Else
                    {
                    	$\textbf{For Each } t \text{ in } Trends$
                    	{
                    	    \If{$ slope\geq S_{t} \text{ and } dur\geq D_{t}$}
                    	    {
                    	        $a[P[len-i ] : P[len]] = 1$  \tcp*{current trend greater than one of the saved trend so mark anomalous}
                    	    }
                    	}
                    }
            }
       	     $i = i+1$	\;
       }
       \KwRet $a$ \tcp*{list with 0s and anomalous indexes represented by 1}
  } 
\caption{Precog Algorithm}\label{alg:precog}
\end{algorithm}

%\subsection{Precog with Maximum Filtration (PrecogMF)}
%It was found that after running the Precog on a sample dataset for evaluation, there were cases where there exists an increasing trend in the new data but the overall maximum of this trend is less than the training data. As a result, these trends were getting reported as anomalous but should not have been as the training data has seen a similar trend with a higher value observation hence this increasing trend is normal and can be ignored. 

%Thus, a modification to Precog is introduced called Precog with Maximum Filtration to check if the reported anomalous window's maximum value is less than the one found in the training data. If it is found to be less, then the window is declared as normal otherwise, not. Thus further increasing the accuracy of the algorithm. 
\section{Evaluation}
We design experiments to answer the questions:
\begin{itemize}
    \item \textbf{Q1. Memory Leak Detection Accuracy}: how accurate is Precog in the detection of memory leaks?
    \item \textbf{Q2. Scalability}: How does the algorithm scale with the increase in the data points?
    \item \textbf{Q3. Parameter Sensitivity}: How sensitive is the algorithm when the parameters values are changed?
\end{itemize}

We have used F1-Score (denoted as F1) to evaluate the performance of the algorithms. Evaluation tests have been executed on a machine with 4 physical cores (3.6 GHz Intel Core i7-4790 CPU) with hyperthreading enabled  and 16 GB of RAM. These conditions are similar to a typical cloud VM. It is to be noted that the algorithm detects the cases where there is an ongoing memory leak and assumes that previously there was no memory leak. For our experiments, hyper-parameters are set as follows. The maximum threshold $U$ is set to 100 and the defined critical time $C$ is set to 7 days. The smoothing window size is 1 hour and re-sampling time resolution was set to 5 minutes. Lastly, the minimum R-squared score $R2_{min}$ for a line to be recognized as a good fit is set to 0.75.  65\% of data was used for training and the rest for testing. However, we also show experiments on parameter sensitivity in this section.

\subsection{Q1. Memory Leak Detection Accuracy}
To demonstrate the effectiveness of the developed algorithm, we initially synthetically generated the timeseries. Table~\ref{tab1:synthetic} shows the F1 score corresponding to each memory leak pattern and also the overall F1 score. 
\begin{table}[htbp]
\centering
\caption{Synthetically generated timeseries corresponding to each memory leak pattern and their accuracy score.}\label{tab1:synthetic}
\begin{tabular}{|l|l|l|l|l|l|}
\hline
\textbf{Memory Leak Pattern} &  \textbf{+ve cases}&  \textbf{-ve cases}&  \textbf{F1 Score} &\textbf{Recall} & \textbf{Precision} \\
\hline
Linearly Increasing & 30 & 30 & 0.933 &0.933 & 0.933 \\
Linearly Increasing(with Noise)  & 30 & 30 & 0.895 &1.0 &0.810\\

Sawtooth & 30 & 30 & 0.830 &0.73 &0.956\\
\hline
Overall & 90 & 90 & 0.9 & 0.9 & 0.91\\
\hline
\end{tabular}
\end{table}
Table~\ref{tab1:synthetic} shows that Precog is able to reach an overall accuracy of 90\%. 

In addition, to demonstrate the effectiveness of the developed algorithm on the real cloud workloads, we evaluated Precog on the real Cloud dataset provided by Huawei Munich which consists of manually labeled memory leak data from 60 VMs spanned over 5 days and each time series consists of an observation every minute. Out of these 60 VMs, 20 VMs had a memory leak. Such high number of VMs having memory leaks is due to the fact that applications with memory leak were deliberately run on the infrastructure. The algorithm achieved the F1-Score of \texttt{0.857}, recall equals to \texttt{0.75} and precision as \texttt{1.0}. Average prediction time per test data containing approximately \texttt{500} points  is \texttt{0.32} seconds.

\begin{figure*}[t] 
\begin{subfigure}{.48\textwidth}
  \centering
  \includegraphics[width=0.9\linewidth]{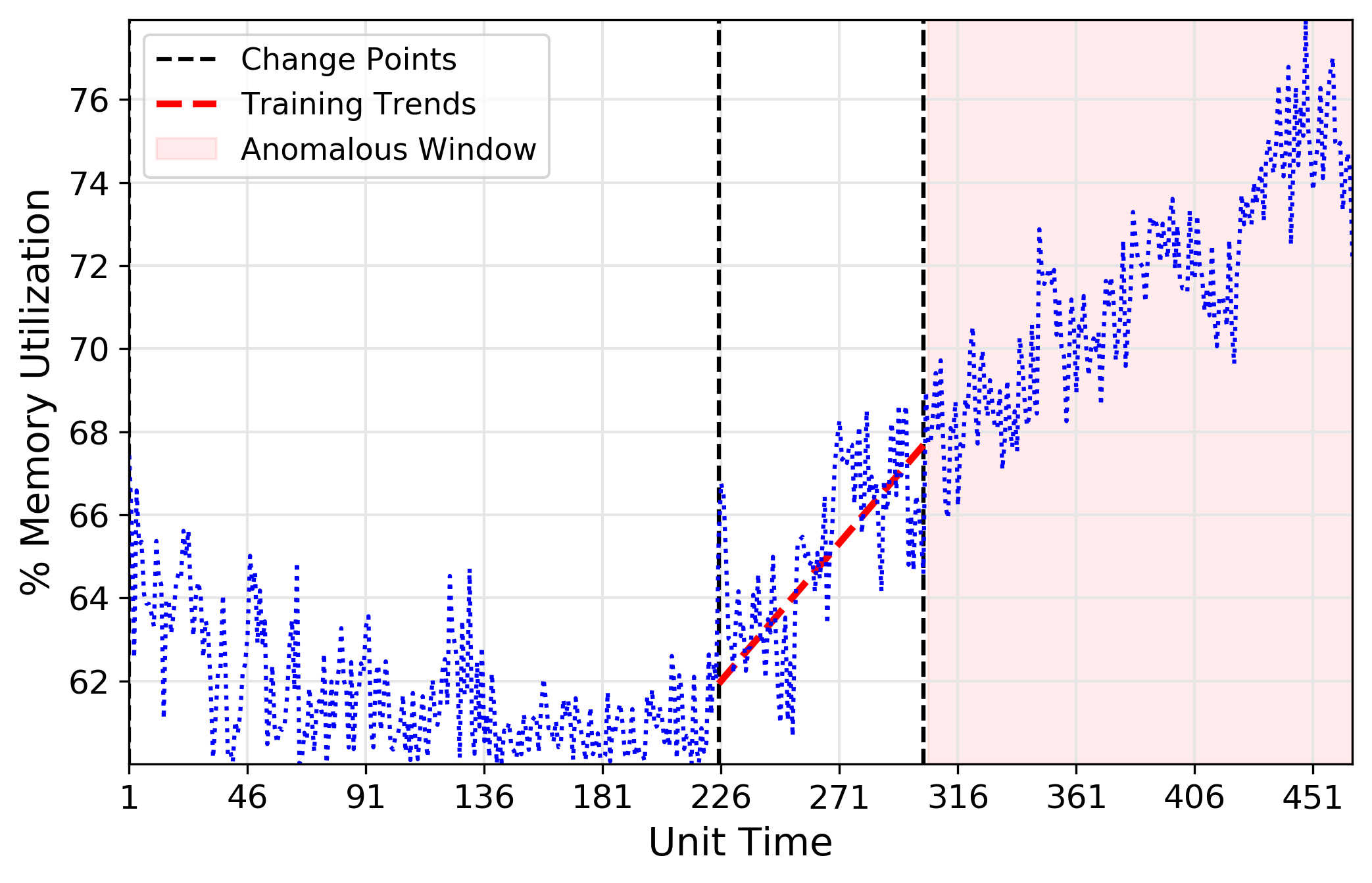}
    \captionof{figure}{Linearly increasing}
  \label{fig:lbr_perf_linear_increase}
\end{subfigure}%
\begin{subfigure}{0.48\textwidth}
  \centering
  \includegraphics[width=0.9\linewidth]{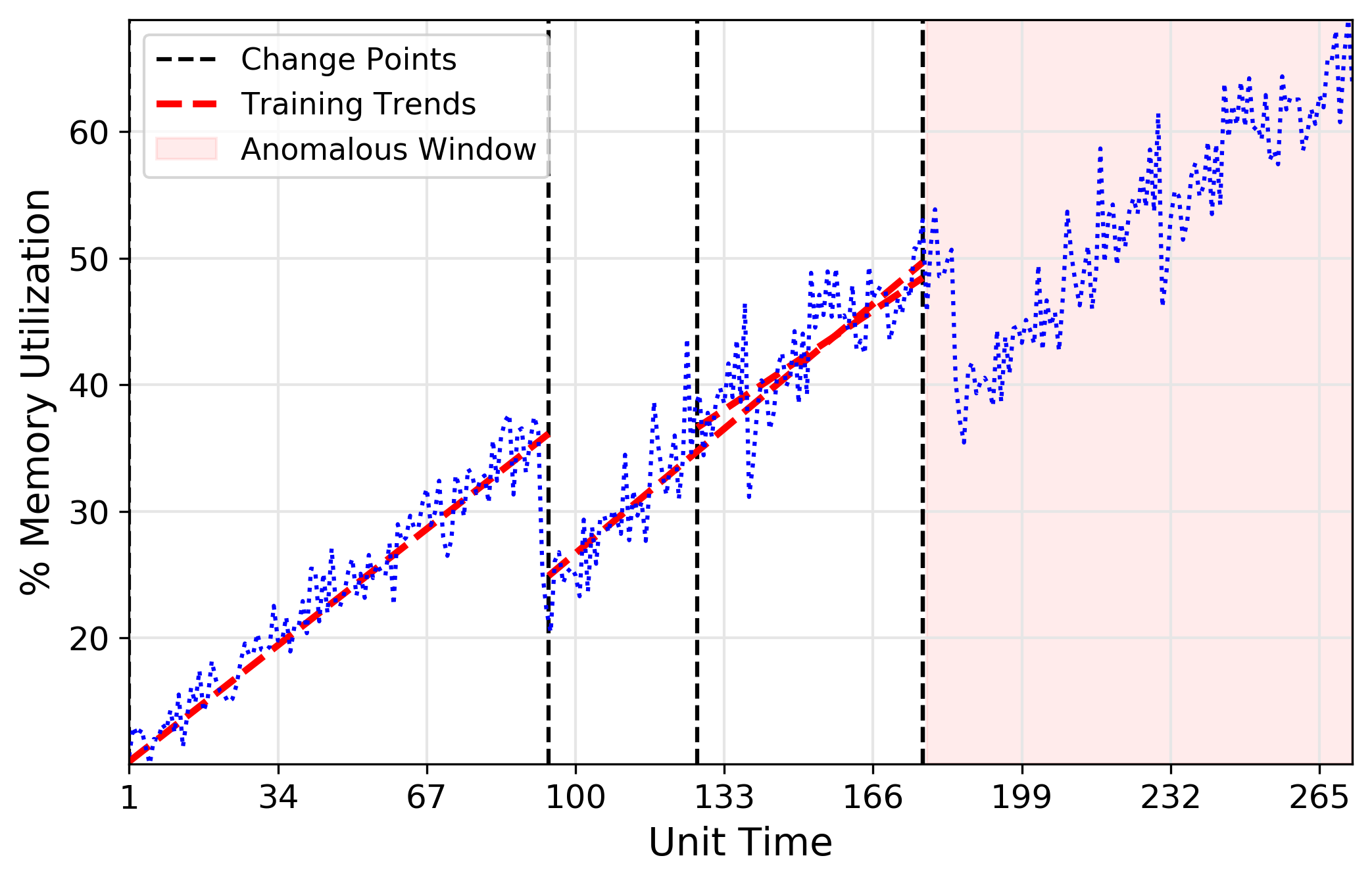}
      \captionof{figure}{Sawtooth linearly increasing }
  \label{fig:lbr_perf_saw_tooth_increase} 
\end{subfigure}
\begin{subfigure}{.48\textwidth}
  \centering
  \includegraphics[width=0.9\linewidth]{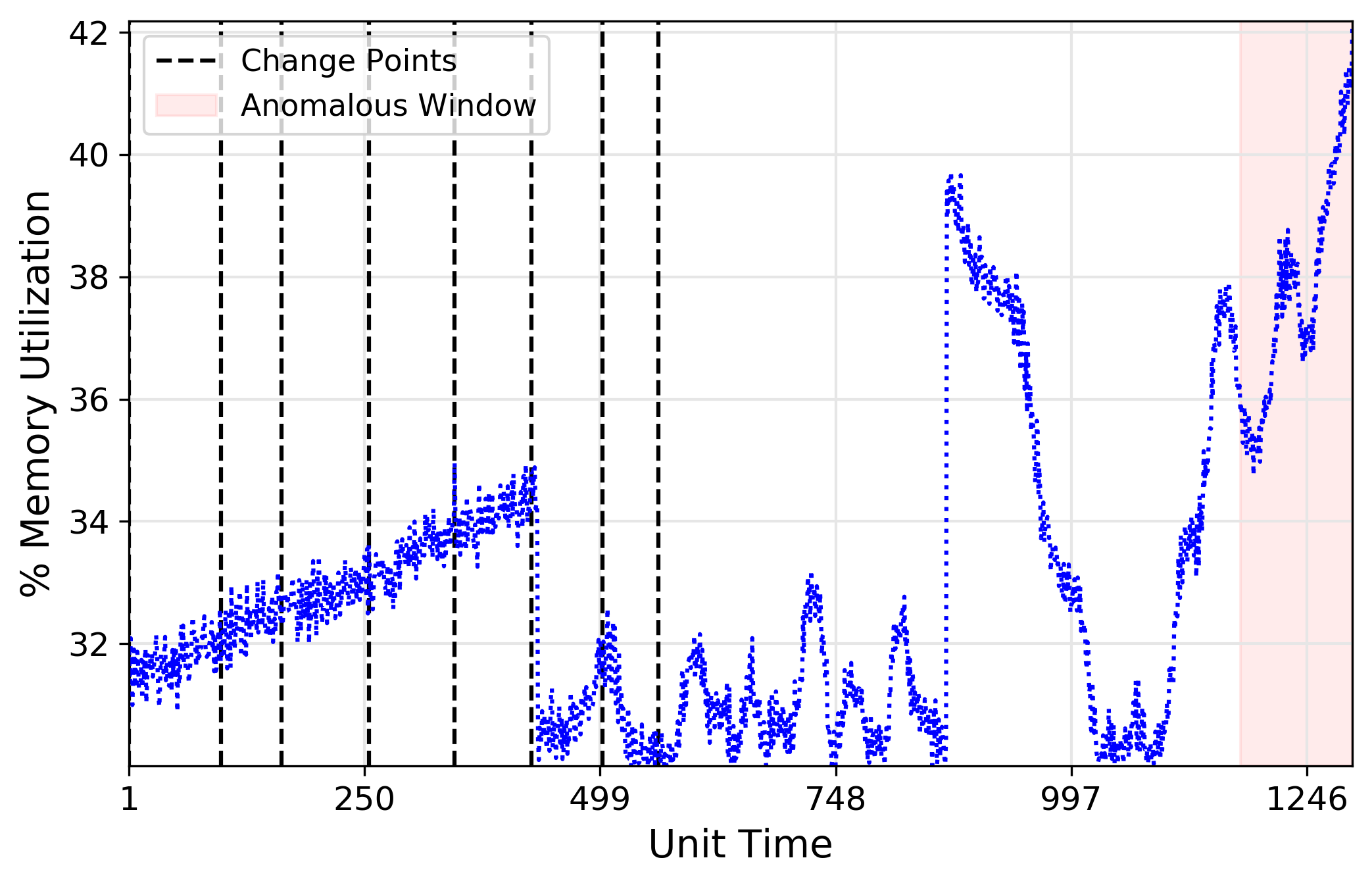}
    \captionof{figure}{Linearly increasing without trends \\detected in training data }
  \label{fig:lbr_perf_no_training_trends} 
\end{subfigure}%%
\begin{subfigure}{0.48\textwidth}
  \centering
  \includegraphics[width=0.9\linewidth]{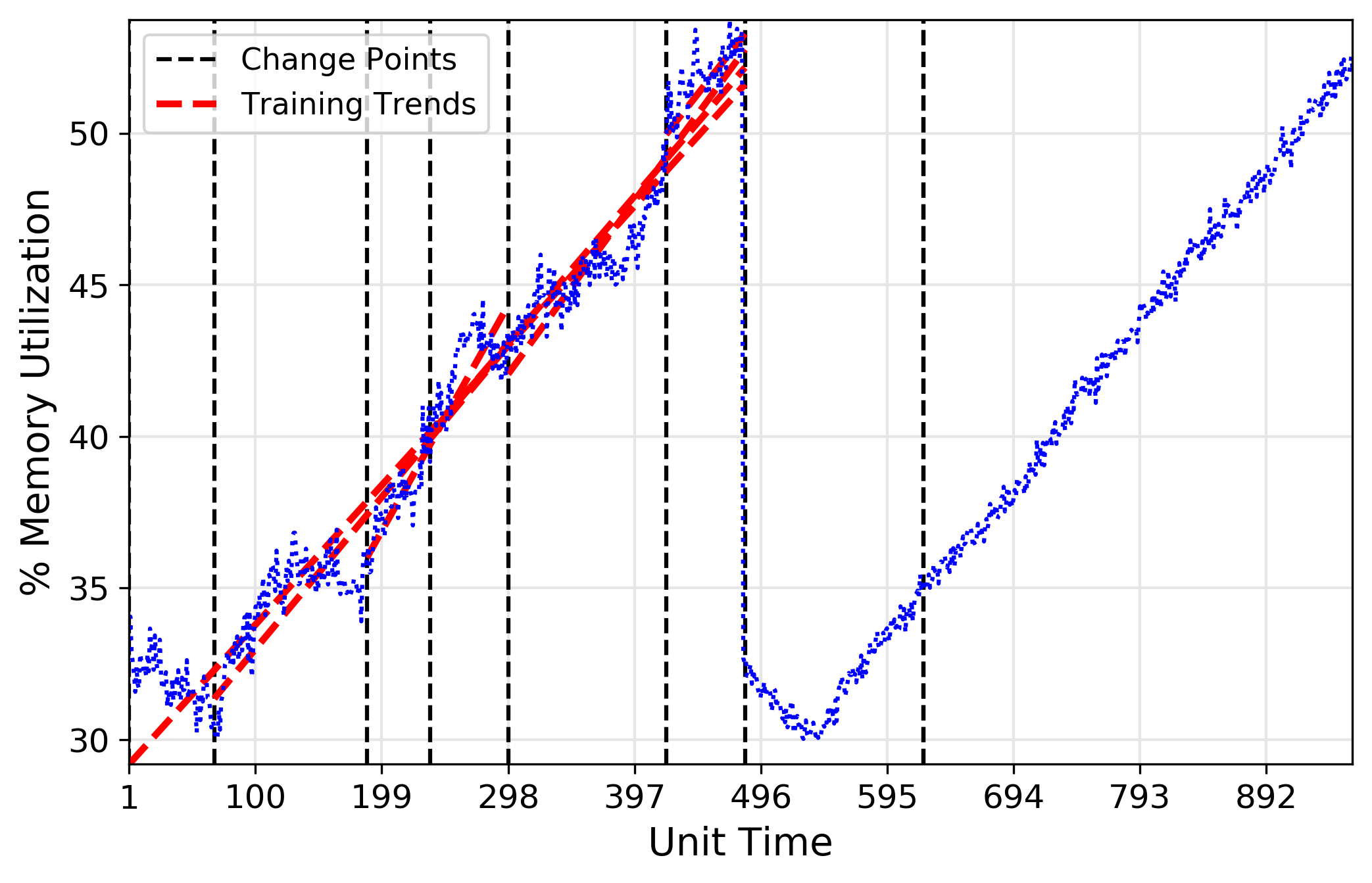}
    \captionof{figure}{Linearly increasing with similar trend as training data and correctly not detected }
  \label{fig:lbr_perf_no_memory_leak} 
\end{subfigure}
 \caption{Algorithm result on 3 difficult cases having memory leak (a-c) and one case not having a memory leak (d).}
  \label{algorithms_test_results}
\end{figure*}

Furthermore, we present the detailed results of the  algorithm on the selected 4 cases shown in the Figure~\ref{algorithms_test_results} : simple linearly increasing memory utilization, sawtooth linearly increasing pattern, linearly increasing pattern with no trends detected in training data, and linearly increasing with similar trend as training data. The figure also shows the change points, training trends and the detected anomalous memory leak window for each of the cases.

For the first case shown in Fig.~\ref{fig:lbr_perf_linear_increase}, memory utilization is being used normally until it suddenly starts to increase linearly. The algorithm detected one training trend and reported the complete test set as anomalous.  The test set trend is having similar slope as training trend but with a longer duration and higher memory usage hence it is reported as anomalous. 
    
 In the second case (Fig.~\ref{fig:lbr_perf_saw_tooth_increase}), the trend represents commonly memory leak sawtooth pattern where the memory utilization increases upto a certain point and then decreases (but not completely zero) and then again it start to increase in the similar manner. The algorithm detected three training trends and reported most of the test set as anomalous. The test set follows a similar trend as captured during the training but with the higher memory utilization, hence it is reported. 
    
In the third case (Fig.~\ref{fig:lbr_perf_no_training_trends}), no appropriate training trend was detected in the complete training data but, the algorithm is able to detect an increasing memory utilization trend in the test dataset.
    
In Fig.~\ref{fig:lbr_perf_no_memory_leak}, the VM does not have a memory leak but its memory utilization was steadily increasing which if observed without the historic data seems to be a memory leak pattern. However, in the historic data, the same trend is already observed and therefore it is a normal memory utilization pattern. Precog using the historic data for detecting the training trends and then comparing them with the test data correctly reports that trend as normal and hence does not flag the window as anomalous.  It is also to be noted that, if the new data's maximum goes beyond the maximum in the training data with the similar trend then it will be regarded as a memory leak. 

\subsection{ Q2. Scalability}
\begin{figure*}[t] 
\begin{subfigure}{.48\textwidth}
  \centering
  \includegraphics[width=0.845\linewidth]{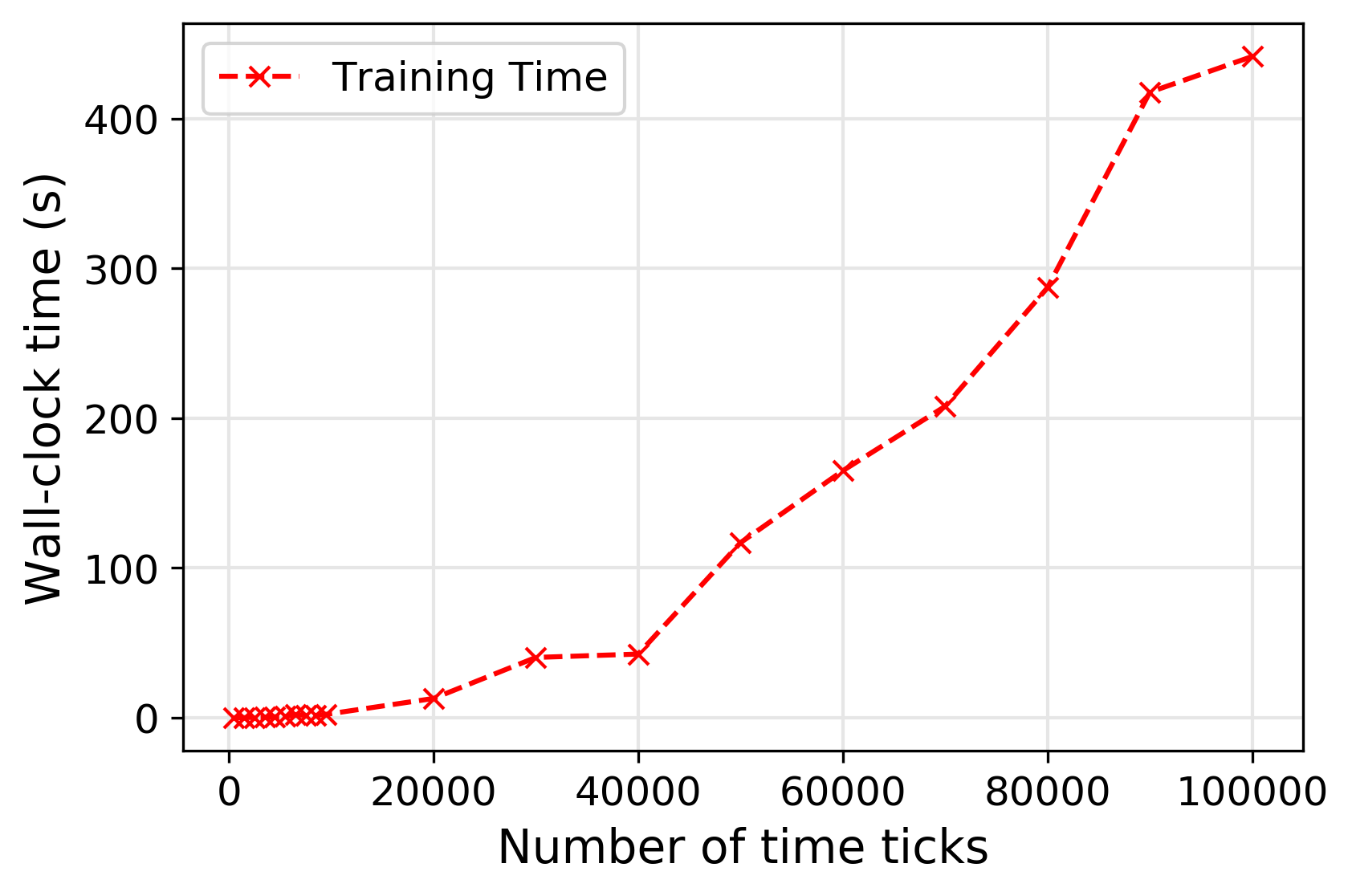}
    \captionof{figure}{Training Time}
  \label{fig:train_time}
\end{subfigure}%
\begin{subfigure}{0.48\textwidth}
  \centering
  \includegraphics[width=0.85\linewidth]{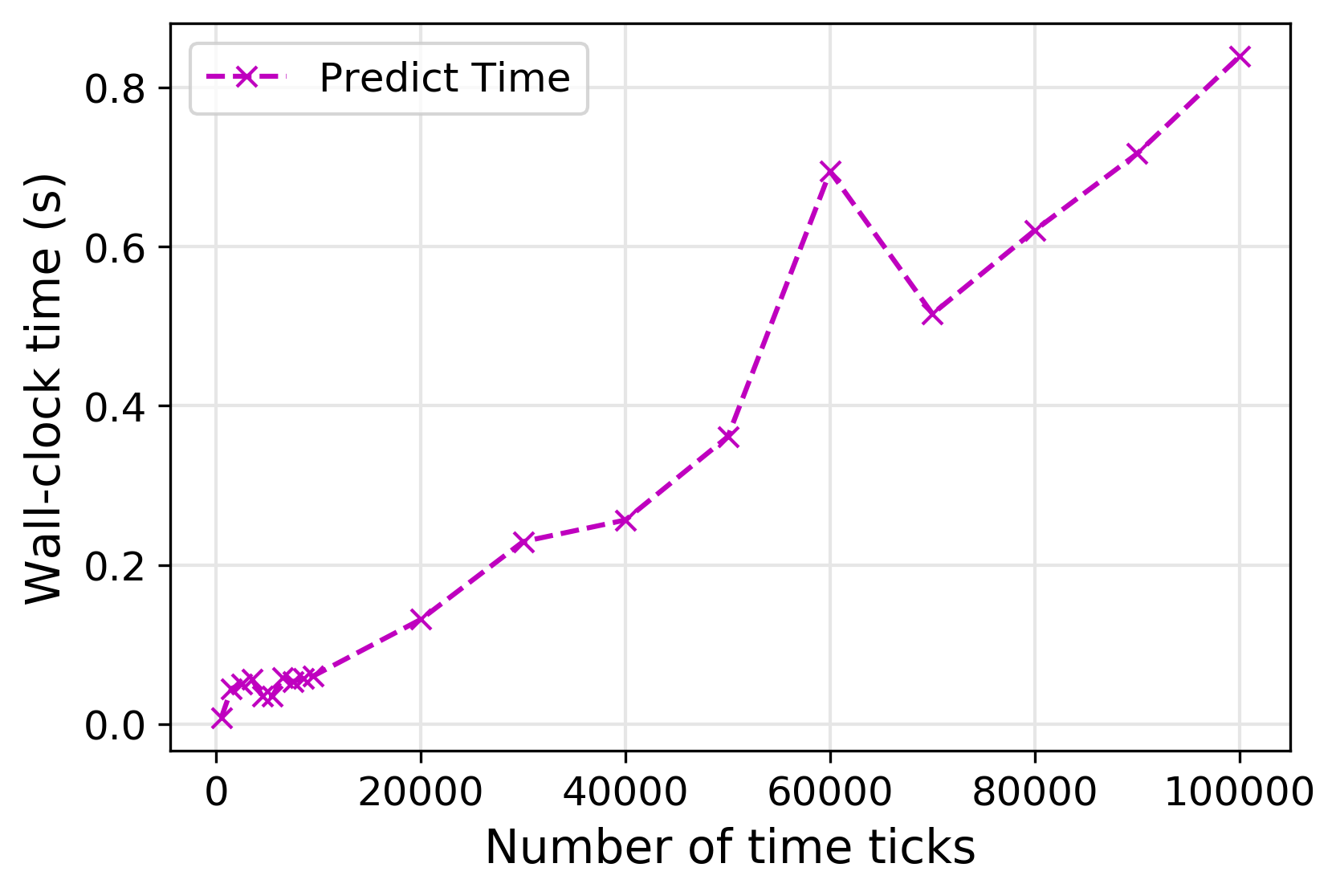}
      \captionof{figure}{Prediction Time }
  \label{fig:predict_time} 
\end{subfigure}
 \caption{Precog's prediction method scale linearly.}
  \label{algorithms_time_results}
\end{figure*}

Next, we verify that our prediction method scale linearly. We repeatedly duplicate our dataset in time ticks, add Gaussian noise. Figure~\ref{fig:predict_time} shows that Precog' predict method scale linearly in time ticks. Precog does provide the prediction results under 1 second for the data with 100,000 time ticks. However, the training method shown in Figure~\ref{fig:train_time} is quadratic in nature but training needs to conducted once a week or a month and it can be done offline as well.

\subsection{ Q3. Parameter Sensitivity}
\begin{figure*}[t] 
\begin{subfigure}{.48\textwidth}
  \centering
  \includegraphics[width=0.85\linewidth]{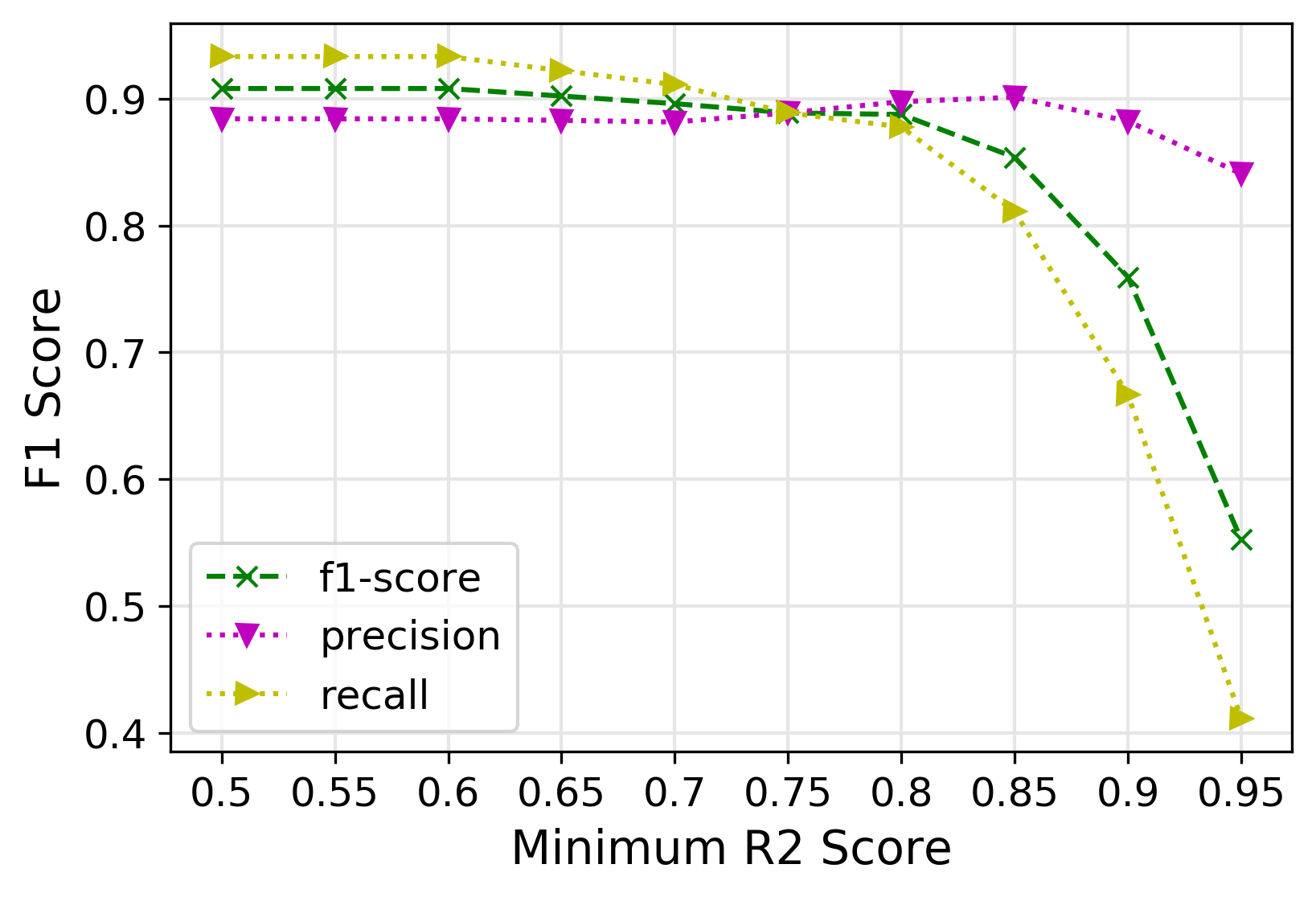}
  \label{fig:parameter_r2_score}
\end{subfigure}%
\begin{subfigure}{0.48\textwidth}
  \centering
  \includegraphics[width=0.85\linewidth]{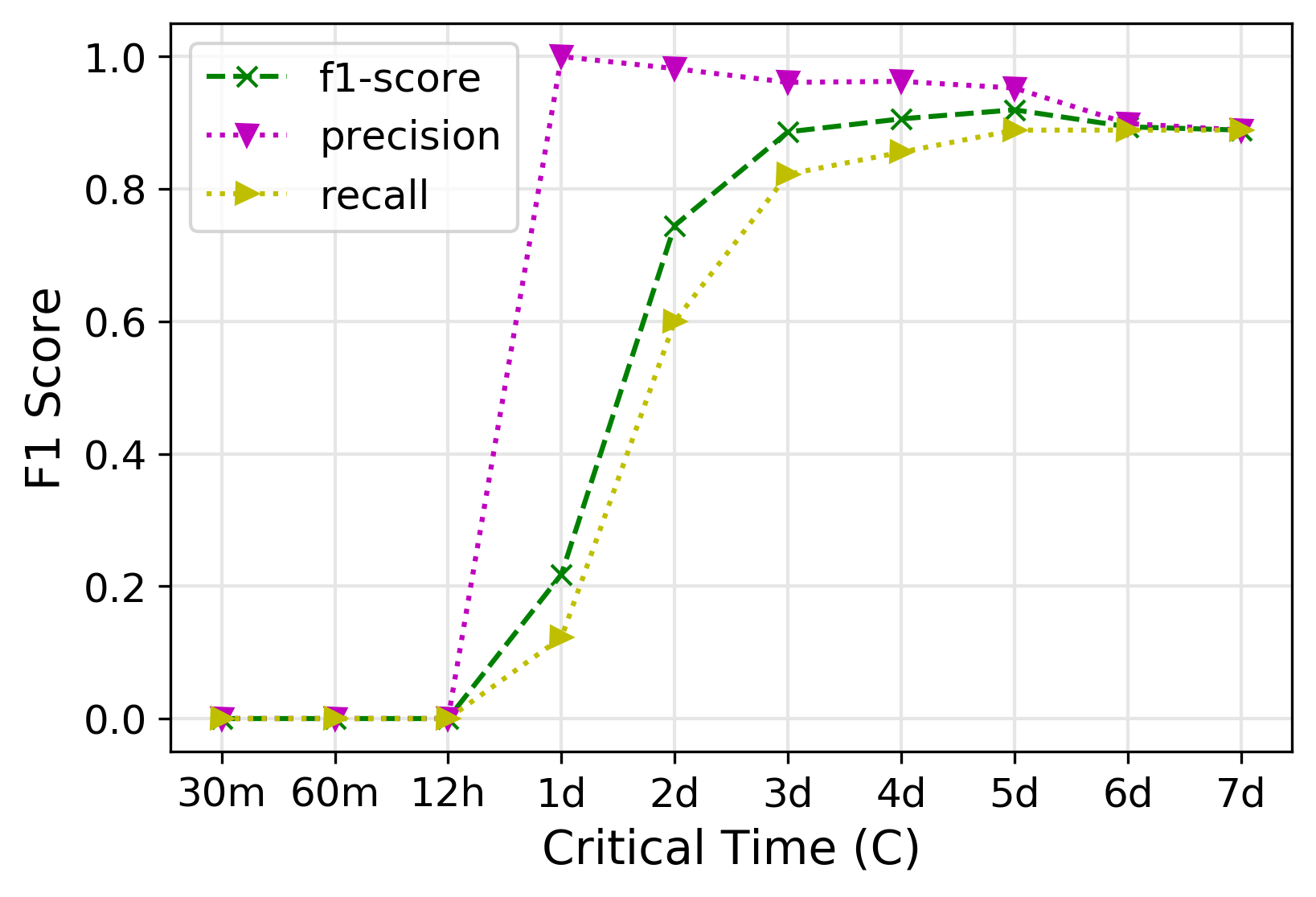}
  \label{fig:parameter_cric_time} 
\end{subfigure}
 \caption{Insensitive to parameters: Precog
performs consistently across parameter values.}
  \label{algorithms_parameter_results}
\end{figure*}

Precog requires tuning of certain hyper-parameters like R2 score, and critical time, which currently are set manually based on the experts knowledge. Figure~\ref{algorithms_parameter_results} compares performance for different parameter values, on synthetically generated dataset.  Our algorithm perform consistently well across values. Setting minimum R2 score above 0.8 corresponds to stricter fitting of the line and that is why the accuracy drops. On the other hand, our data mostly contains trend lines which would reach threshold withing 3 to 4 days, therefore setting minimum critical time too less (less than 3 days) would mean the trend line never reaching threshold within the time frame and hence decreasing the accuracy. These experiments shows that these parameters does play a role in the overall accuracy of the algorithm but at most of the values algorithm is insensitive to them. Furthermore, to determine these automatically based on the historic data is under progress and is out of the scope of this paper. 

\section{Conclusion}
Memory leak detection has been a research topic for more than a decade. Many approaches have been proposed to detect memory leaks, with most of them looking at the internals of the application or the object's allocation and deallocation. The Precog algorithm for memory leak detection presented in the current work is most relevant for the cloud-based infrastructure where cloud administrator does not have access to the source code or know about the internals of the deployed applications.  The performance evaluation results showed that the Precog is able to achieve a F1-Score of 0.85 with less than half a second prediction time on the real workloads. This algorithm can also be useful in the Serverless Computing where if a function is leaking a memory then its successive function invocations will add on to that and resulting in a bigger memory leak on the underneath system. Precog running on the underneath system can detect such a case. 

Prospective directions of future work include developing online learning-based approaches for detection and as well using other metrics like CPU, network and storage utilization for further enhancing the accuracy of the algorithms and providing higher confidence in the detection results.

\section*{ACKNOWLEDGEMENTS}
This work was supported by the funding of the German Federal Ministry of Education and Research (BMBF) in the scope of the Software Campus program.  The authors also thank the anonymous reviewers whose comments helped in improving this paper. 


\bibliographystyle{splncs04}
\bibliography{bib}
\end{document}